\documentstyle[12pt]{article}
\begin{document}
\pagestyle{empty}

\hspace*{12.4cm}IU-MSTP/23 \\
\hspace*{13cm}hep-th/9707229 \\
\hspace*{13cm}July, 1997

\begin{center}
 {\Large\bf On the Three-Point Couplings \\
 in Toda Field Theory}
\end{center}

\vspace*{1cm}
\def\thefootnote{\fnsymbol{footnote}}
\begin{center}
{\sc Hiroshi Igarashi,}$^1$
{\sc Yoshio Takimoto}$^1$ and {\sc Takanori Fujiwara}$^2$
\end{center}
\vspace*{0.2cm}
\begin{center}
{\em $\ ^{1}$ Graduate School of Science and Engineering,
Ibaraki University, Mito 310, Japan}\\
{\em $\ ^{2}$ Department of Physics, Ibaraki University,
Mito 310, Japan}\\
\end{center}
\vfill
\begin{center}
{\large\sc Abstract}
\end{center}
Correlation functions of Toda field vertices are investigated 
by applying the method of integrating zero-mode developed for Liouville 
theory. We generalize the relations among the zero-, two- and 
three-point couplings known in Liouville case to arbitrary 
Toda theories. Two- and three-point functions of Toda vertices 
associated with the simple roots are obtained.

\eject
\pagestyle{plain}
Since the impressive success of Goulian and Li \cite{goulian91}
that the three-point correlators of the Liouville vertex functions 
appearing as the dressing factor of minimal model can be obtained 
by continuing the central charge to a 
certain value where the functional integral can be carried out 
exactly via free field technique \cite{gupta}, several authors 
have pushed 
forward their method \cite{francesco,sakai,aoki} and generalized to 
define arbitrary three-point couplings
in Liouville theory  \cite{sakai,dorn94,zamolodchikov}. 
Since Liouville theory is a Toda field 
theory associated with the Lie algebra $sl_2$ and as two dimensional 
field theory Toda theories possess distinguished properties 
such as the $W$ symmetry \cite{wsymm,bilal,mansfield}, it is natural to ask 
whether the similar functional integral approach can apply for 
the general Toda theories. 

In this note we shall investigate correlators of the vertex 
functions of Toda fields by applying the method of ref. 
\cite{gupta,goulian91}. We see that the zero-modes of the Toda fields 
can be integrated 
as in Liouville case and the remaining functional integrations
over nonzero-modes can be carried out via free field technique
if all the $s$-parameters 
are nonnegative integers. We thus arrive at complicated 
multiple integrals over complex planes. Unlike the Liouville case, 
we can not give closed forms of the integrals for three-point couplings. 
This prevent us from getting explicit expressions for general 
three-point couplings. We find, however, some universality among the 
three-point couplings for the vertex functions associated to simple 
roots. The relations among the zero-, two- and 
three-point couplings found in Liouville case \cite{dorn94} can also be 
generalized to Toda theory. In particular, we will give closed 
expressions for two- and three-point functions of the vertices associated 
to simple roots. 

We consider the Toda field theory associated with the simple 
Lie algebra ${\cal G}$ of rank $r$. The Toda field $\varphi$
is an $r$-component vector in the root space. Let us denote 
the simple roots by $\alpha^a$ ($a=1,\cdots,r$), then it is 
described by the classical field equations
\begin{eqnarray}
  \label{cact}
  \partial_z\partial_{\bar z}\varphi-\frac{\mu^2}{8}\sum_{a=1}^r\alpha^a
  {\rm e}^{\alpha^a\cdot\varphi}=0~.
\end{eqnarray}
For ${\cal G}=sl_2$ this reduces to the Liouville equation. 

The equation of motion (\ref{cact}) is invariant under the 
conformal reparametrization
$z\rightarrow\xi=f(z)$ by the shift
\begin{eqnarray}
  \label{transf}
  \varphi \rightarrow\varphi -\rho\ln\partial_zf
  \partial_{\bar z}\bar f ~,
\end{eqnarray}
where $\rho$ is a vector in the root space satisfying 
\begin{eqnarray}
  \label{defrho}
  \rho\cdot \alpha^a=1 ~,
\end{eqnarray}
for any simple root. In terms of the fundamental weights $\lambda^a$ 
($a=1,\cdots,r$) defined by \\ $2\lambda^a\cdot\alpha^b/(\alpha^a)^2
=\delta^{ab}$, it is given by 
\begin{eqnarray}
  \label{rho}
  \rho=\sum_{a=1}^r\frac{2\lambda^a}{(\alpha^a)^2}
\end{eqnarray}

In quantum theory we start from the action
\begin{eqnarray}
  \label{qact}
  S[\hat g;\varphi]=\frac{1}{8\pi}\int d^2z\sqrt{\hat g}\biggl(
  \hat g^{\alpha\beta}\partial_\alpha\varphi\cdot\partial_\beta\varphi
  +Q\cdot\varphi\hat R+\frac{\mu^2}{\gamma^2}
  \sum_{a=1}^r{\rm e}^{\gamma\alpha^a\cdot\varphi}\Biggr) ~,
\end{eqnarray}
where $\hat g_{\alpha\beta}$ is the fiducial metric on the Riemann 
surface. We have introduced couplings $Q$, a vector in the root space, 
with the curvature and $\gamma$ in the Toda potential. They are 
determined by requiring the conformal invarince. 

The central charge of the Toda theory can be found most easily by 
applying the DDK's argument \cite{ddk}. We note that the stress 
tensor for $\mu^2=0$ and in locally flat 
coordinates is given by
\begin{eqnarray}
  \label{st}
  T(z)=-\frac{1}{2}(\partial_z\varphi)^2+\frac{1}{2}Q
  \cdot\partial_z^2\varphi
\end{eqnarray}
Using the operator product relation $\varphi_j(z)\varphi_k(w)
\sim \delta_{jk}\ln|z-w|^2$, one can easily obtain
\begin{eqnarray}
  \label{TT}
  T(z)T(w)\sim\frac{1}{2}\frac{r+3Q^2}{(z-w)^4} ~.
\end{eqnarray}
The central charge of the Toda sector is thus given by 
\begin{eqnarray}
  \label{cc}
  c_\varphi=r+3Q^2 ~.
\end{eqnarray}
One can also determine the conformal dimension of arbitrary Toda 
vertex functions. Let $\beta$ be an arbitrary vector in the root space, 
then the conformal dimension of the vertex function ${\rm e}^{\beta\cdot
\varphi}$ is given by
\begin{eqnarray}
  \label{cdtvf}
  \Delta({\rm e}^{\beta\cdot\varphi})
  =\frac{1}{2}\beta\cdot(Q-\beta) ~.
\end{eqnarray}
In particular the Toda potential in (\ref{qact}) must be a 
(1,1)-conformal field by the requirement of conformal invariance. This 
relates $Q$ with the coupling constant $\gamma$ as 
\begin{equation}
  \label{cdtp}
  \gamma\alpha^a\cdot(Q-\gamma\alpha^a)=2 \qquad 
\end{equation}
for any simple root. 
This is the quantum version of the classical relation (\ref{defrho})
and has already been obtained in ref. \cite{hollowood89}.
Since the $r$ simple roots are linearly independent each other, one 
can uniquely find the expression for $Q$ as 
\begin{equation}
  \label{Q}
  Q=2\Biggl(\frac{1}{\gamma}\rho+\gamma\bar\rho\Biggr) ~,
\end{equation}
where $\bar\rho$ is simply the sum of all fundamental weights, {\it i.e.}, 
\begin{equation}
  \label{rhoast}
  \bar\rho=\sum_{a=1}^r\lambda^a~.
\end{equation}
For simply-laced Lie algebra one has $\bar\rho=\rho$. Hence, the 
renormalization of the couplings $Q$ is simply a rescaling of the 
classical value as observed in Liouville theory. 
Since simple roots with different lengths coexist for nonsimply-laced 
cases, the couplings $Q$ can not be proportional to $\rho$ as can be 
easily seen from (\ref{cdtp}). Using (\ref{rhoast}), one can parametrize 
the central charge (\ref{cc}) in terms of $\gamma$. The Virasoro 
central charge in quantum Toda theories has been obtained in ref. 
\cite{mansfield83}. A complete list for 
any simple Lie algebra is given in ref.\cite{hollowood89}. 

Let us denote by $c_{gh}$ and $c_M$ the central charges of the ghost 
and matter sectors which are conformally coupled to the Toda fields. 
Then the total central charge must vanish to ensure the conformal 
invariance
\begin{equation}
  \label{tcc}
  c_\varphi+c_M+c_{gh}=0~.
\end{equation}
From (\ref{cc}), (\ref{cdtp}) and (\ref{Q}) we obtain the coupling 
constant as
\begin{equation}
  \label{gamma}
  \gamma=\sqrt{\frac{-c_{gh}-c_M-r
      +24(|\rho||\bar\rho|-\rho\cdot\bar\rho)}{48|\bar\rho|^2}}
    -\sqrt{\frac{-c_{gh}-c_M-r
      -24(|\rho||\bar\rho|+\rho\cdot\bar\rho)}{48|\bar\rho|^2}} ~, 
\end{equation}
where we have chosen a branch of the square roots so that the correct 
classical limit is recovered for $c_M\rightarrow-\infty$. Up to trivial 
rescaling of $\gamma$, this reproduces the well-known results for 
Liouville theory \cite{ddk}. 
Due to the extended conformal symmetries \cite{bilal} we must include 
ghosts with 
higher conformal dimensions for general Toda theories. By requiring 
$\gamma$ to be real we find the bound for the matter central charge 
\begin{equation}
  \label{cmbound}
  c_M\le-c_{gh}-r-24(|\rho||\bar\rho|+\rho\cdot\bar\rho)
\end{equation}

To illustrate in a concrete example consider $A_n$ Toda theory. The 
holomorphic ghost sector of this system consists of pairs of ghost 
and anti-ghost with conformal weights $(-j,0)$ and $(j+1,0)$ 
for $j=1,\cdots,n$, each having central charge $-2(6j^2+6j+1)$. The 
total ghost central charge is then given by
\begin{equation}
  \label{tghcc}
  c_{gh}=-2n(2(n+1)^2+2(n+1)+1) ~.
\end{equation}
On the other hand the squared norm of $\rho$ is easily found to be 
\begin{equation}
  \label{norm}
  \rho^2=\frac{1}{12}n(n+1)(n+2) ~.
\end{equation}
Putting these into (\ref{cmbound}) and noting $\bar\rho=\rho$, we obtain 
\begin{equation}
  \label{cmanbound}
  c_M\le n~.
\end{equation}
This generalizes the well-known $c=1$ barrier for Liouville gravity 
to $A_n$ Toda gravity. 

We now turn to an arbitrary $N$-point functions of Toda vertices on the 
surface of spherical topology. It is defined by 
\begin{eqnarray}
  \label{N-ptfunc}
  G_N(z_1,\cdots,z_N|\beta_1,\cdots,\beta_N)
  =\int{\cal D}_{\hat g}\varphi{\rm e}^{-S[\hat g,\varphi]}
  \prod_{j=1}^{N}{\rm e}^{\beta_j\cdot\varphi(z_j)}~.
\end{eqnarray}
We decompose the Toda field into 
the zero-mode $\varphi_0$ and the nonzero-mode $\tilde\varphi$ by
\begin{eqnarray}
  \label{zero-mode}
  \varphi=\varphi_0+\tilde\varphi ~, \qquad 
  \varphi_0=\frac{\int d^2z\sqrt{\hat g}\varphi}{\int d^2z\sqrt{\hat g}} ~.
\end{eqnarray}
Then the integrations over the zero-mode can be carried out as in the 
Liouville case. We thus arrive at the functional integral
\begin{eqnarray}
  \label{pathint1}
   G_N(z_1,\cdots,z_N|\beta_1,\cdots,\beta_N)
  =|\det\alpha|^{-1}
  \prod_{a=1}^r \Biggl(\frac{\mu^2}{8\pi\gamma^2}\Biggr)^{s_N^a}
  \frac{\Gamma(-s_N^a)}{\gamma}
  I_N(z_1,\cdots,z_N|\beta_1,\cdots,\beta_N)~,
\end{eqnarray}
where $\alpha$ is an $r\times r$ matrix with $\alpha^a$ as the $a$-th 
row elements and $s_N^a$ is defined by
\begin{eqnarray}
  \label{sna}
  s_N^a=\frac{2\lambda^a\cdot\Biggl(Q
    -\displaystyle{\sum_{j=1}^N}\beta_j\Biggr)}{\gamma(\alpha^a)^2} ~.
\end{eqnarray}
The $N$-point function $I_N$ in the rhs of (\ref{pathint1}) is given by
\begin{eqnarray}
  \label{I}
  I_N(z_1,\cdots,z_N|\beta_1,\cdots,\beta_N)=
  \int{\cal D}_{\hat g}\tilde\varphi{\rm e}^{-S_0[\hat g;\tilde\varphi]}
  \prod_{a=1}^r
  \Biggl(\int d^2z\sqrt{\hat g}{\rm e}^{\gamma\alpha^a\cdot\tilde\varphi}
  \biggr)^{s_N^a}
  \prod_{j=1}^{N}{\rm e}^{\beta_j\cdot\tilde\varphi(z_j)} ~.
\end{eqnarray}
where $S_0[\hat g;\tilde\varphi]$ is the action of $r$ free massless 
bosons. 

In general $s_N^a$ take generic values for arbitrary $\beta$'s 
and the functional 
integration (\ref{I}) can not be carried out explicitly. 
We avoid this difficulty by considering all the $s_N^a$ being 
nonnegative integers and then continuing them back to the original generic 
values. In the case of Liouville theory $s$ can be made a nonnegative 
integer by adjusting the central charge of the matter sector. 
Furthermore, continuation back to generic $s$ is known 
by explict construction \cite{dorn94,zamolodchikov}. 
As for the Toda theory, all the $s_N^a$, 
in general, can not be made nonnegative integers at the same time 
by varying $c_M$. We can acheive this by adjusting one of the 
$\beta$'s for $N\ge1$. In the following argument, 
we simply assume that the continuation back to arbitrary $s_N^a$ 
exists. We shall see, however, that the existence of such continuation 
procedure 
is not so essential in the present work. It is only necessary to 
show explicitly the conformality of (\ref{I}) under M\"obius 
transformations. 

We choose the fiducial metric $\hat g_{\alpha\beta}$ to be everywhere 
flat except at 
infinity on the complex plane. In this case one can perform the 
functional integration as 
\begin{eqnarray}
  \label{pathint2}
  I_N(z_1,\cdots,z_N|\beta_1,\cdots,\beta_N)&=&
  \int{\cal D}_{\hat g}\tilde\varphi{\rm e}^{-S_0[\hat g;\tilde\varphi]}
  \int\prod_{a=1}^r\prod_{I=1}^{s_N^a}d^2w_I^a
  \sqrt{\hat g(w_I^a)}{\rm e}^{\gamma\alpha^a\cdot\tilde\varphi(w_I^a)}
  \prod_{j=1}^{N}{\rm e}^{\beta_j\cdot\tilde\varphi(z_j)} \nonumber\\
  &=&\int\prod_{a=1}^r\prod_{I=1}^{s_N^a}d^2w_I^a 
  \prod_{i<j}^N|z_i-z_j|^{-2\beta_i\cdot\beta_j}
  \prod_{i=1}^N\prod_{a=1}^r\prod_{I=1}^{s_N^a}|z_i-w_I^a|^{-2\gamma
    \beta_i\cdot\alpha^a} \nonumber\\
  &&\times\prod_{a<b}^N\prod_{I=1}^{s_N^a}\prod_{J=1}^{s_N^b}
  |w_I^a-w_J^b|^{-2\gamma^2\alpha^a\cdot\alpha^b} 
  \prod_{a=1}^r\prod_{I<J}^{s_N^a}
  |w_I^a-w_J^a|^{-2\gamma^2(\alpha^a)^2}~.
\end{eqnarray}
Using the integral representation, one can check explicitly the conformal 
property of (\ref{I}) under the M\"obius transformations 
\begin{eqnarray}
  \label{mobius}
  z=\frac{a\xi+b}{c\xi+d} \qquad  
  (ad-bc=1~, \quad a,b,c,d\in {\bf C})~.
\end{eqnarray}
We see that $I_N$ is transformed as 
\begin{eqnarray}
  \label{i'str}
  I_N(z_1,\cdots,z_N|\beta_1,\cdots,\beta_N)=\prod_{i=1}^N
  |c\xi_i+d|^{4\Delta_i}I_N(\xi_1,\cdots,\xi_N|\beta_1,\cdots,\beta_N) ~,
\end{eqnarray}
where $\Delta_j$ is the conformal dimension of the vertex function 
${\rm e}^{\beta_j\cdot\varphi}$.
In showing this resutl, use has been made of (\ref{cdtp}) and (\ref{sna}). 

We can not determine the $z$-depedence of the $N$-point function 
only from (\ref{i'str}) for $N\ge4$. For 3-point function, we can 
extract the usual conformal structure by fixing $\xi_1=0$, $\xi_2=1$, $\xi_3=\infty$ as
\begin{eqnarray}
  \label{3-ptfunc}
  I_3(z_1,z_2,z_3|\beta_1,\beta_2,\beta_3)
  =\frac{C_3(\beta_1,\beta_2,\beta_3)}{|z_{12}|^{2\Delta_{12}}
    |z_{23}|^{2\Delta_{23}}|z_{31}|^{2\Delta_{31}}}~,
\end{eqnarray}
where we have introduced $z_{ij}=z_i-z_j$ and 
$\Delta_{ij}=\Delta_i+\Delta_j-\Delta_k$ 
for $i,j,k=1,2,3$ (cyclic), and $C_3(\beta_1,\beta_2,\beta_3)$ is defined 
by
\begin{eqnarray}
  \label{C}
  C_3(\beta_1,\beta_2,\beta_3)&=&\lim_{\xi_3\rightarrow\infty}
  |\xi_3|^{4\Delta_3}I_3(0,1,\xi_3|\beta_1,\beta_2,\beta_3) \nonumber\\
  &=&\int\prod_{a=1}^r\prod_{I=1}^{s_3^a}d^2w_I^a
  |w_I^a|^{-2\gamma\beta_1\cdot\alpha^a}
  |1-w_I^a|^{-2\gamma\beta_2\cdot\alpha^a} \nonumber\\
  &&\times\prod_{a=1}^r\prod_{I<J}^{s_3^a}
  |w_I^a-w_J^a|^{-2\gamma^2(\alpha^a)^2}
  \prod_{a<b}^r\prod_{I=1}^{s_3^a}\prod_{J=1}^{s_3^b}
  |w_I^a-w_J^b|^{-2\gamma^2\alpha^a\cdot\alpha^b}
\end{eqnarray}
For the Liouville case the rhs of this expression reduces to 
the integral investigated by Dotsenko and Fateev \cite{dotsenko} 
and be carried out explicity. 
Furthermore, the 
resulting expression for $C_3$ can be continued to arbitrary 
$\beta$'s \cite{dorn94,zamolodchikov}. 
Unfortunately, the closed form of the integral is not 
known to the present authors for Toda theories with $r\ge2$ and we 
can not give explicit extension of the three-point coupling proposed 
in refs. \cite{dorn94,zamolodchikov}. Below we will investigate the relations among the zero-, 
two- and three-point couplings by assuming the existence of  
$C_3(\beta_1,\beta_2,\beta_3)$ for arbitrary $\beta$'s that 
coincides to (\ref{C}) if all the $s_3^a$ are nonnegative integers. 
We will find some nontrivial constraints satisfied by the three-point 
couplings, which arise for $r>1$ and have not appeared in the previous 
studies for the Liouville theory. 

We define the three-point coupling $A_3(\beta_1,\beta_2,\beta_3)$ by
\begin{eqnarray}
  \label{A3}
  G_3(z_1,z_2,z_3|\beta_1,\beta_2,\beta_3)
  =\frac{A_3(\beta_1,\beta_2,\beta_3)}{|z_{12}|^{2\Delta_{12}}
    |z_{23}|^{2\Delta_{23}}|z_{31}|^{2\Delta_{31}}}~.
\end{eqnarray}
From (\ref{pathint1}) and (\ref{3-ptfunc}), the three-point coupling 
$A_3$ can be written as
\begin{eqnarray}
  \label{3ptc}
  A_3(\beta_1,\beta_2,\beta_3)=|\det\alpha|^{-1}
  C_3(\beta_1,\beta_2,\beta_3)
  \prod_{a=1}^r \Biggl(\frac{\mu^2}{8\pi\gamma^2}\Biggr)^{s_3^a}
  \frac{\Gamma(-s_3^a)}{\gamma} ~.
\end{eqnarray}
The two-point function
\begin{eqnarray}
  \label{2ptf0}
  G_2(z_1,z_2|\beta,\beta)=\frac{1}{V_2}
  \int{\cal D}_{\hat g}\varphi{\rm e}^{-S[\hat g;\varphi]}
  {\rm e}^{\beta\cdot\varphi(z_1)}{\rm e}^{\beta\cdot\varphi(z_2)}
\end{eqnarray}
can be obtained from the three-point function by the formula \cite{dorn94}
\begin{eqnarray}
  \label{2ptf}
  G_2(z_1,z_2|\beta,\beta)=-\frac{\mu^2}{8\pi\gamma^2s_2^a}\frac{1}{V_2}
  \int d^2z_3G_3(z_1,z_2,z_3|\beta,\beta,\gamma\alpha^a) 
\end{eqnarray}
where $\gamma s_2^a=2\lambda^a\cdot(Q-2\beta)/(\alpha^a)^2$
and $\alpha^a$ is an arbitrary simple root. $V_2$ 
is the infinite volume of the subgroup of the 
M\"obius transformations (\ref{mobius}) that leave the two points 
$z_1$ and $z_2$ fixed. It is given by
\begin{eqnarray}
  \label{vckv2}
  V_2=\int \frac{d^2z_3|z_{12}|^2}{|z_{23}|^2|z_{31}|^2}~.
\end{eqnarray}
The same volume factor will factorize from the integral (\ref{2ptf}) 
and be canceled by the prefactor. 
We thus obtain the two-point function
\begin{eqnarray}
  \label{2ptf2}
  G_2(z_1,z_2|\beta,\beta)&=&\frac{A_2(\beta)}{|z_{12}|^{4\Delta_\beta}} ~,
  \nonumber\\
  A_2(\beta)&=&|\det\alpha|^{-1}C_3(\beta,\beta,\gamma\alpha^a)
  \prod_{b=1}^r \Biggl(\frac{\mu^2}{8\pi\gamma^2}\Biggr)^{s_2^b}
  \frac{\Gamma(-s_2^b)}{\gamma} ~, 
\end{eqnarray}
where $\Delta_\beta=\beta\cdot(Q-\beta)/2$. Using the 
general expression for the three-point coupling (\ref{3ptc}), we obtain 
\begin{eqnarray}
  \label{2-3}
  A_2(\beta)=-\frac{\mu^2}{8\pi\gamma^2s_2^a}A_3(\beta,\beta,
  \gamma\alpha^a) \qquad (a=1,\cdots,r)~.
\end{eqnarray}
This relation has already been noted for Liouville theory in 
refs. \cite{dorn94}. 
In the case of Toda theory new situation arises. Since  we may 
choose one of the simple roots arbitrarily, (\ref{2ptf2}) implies 
that $C_3(\beta,\beta,\gamma\alpha^a)$ is independent of the simple 
roots $\alpha^a$. 

The partition function $A_0$ can be obtained from the three-point 
function in the manner similar to the two-point function. Let 
$\alpha^{a,b,c}$ be arbitrary simple roots, then 
the partition function is given by 
\begin{eqnarray}
  \label{0ptf}
  A_0&=&|\det\alpha|^{-1}C_3(\gamma\alpha^a,\gamma\alpha^b,\gamma\alpha^c)
  \prod_{d=1}^r\Biggl(\frac{\mu^2}{8\pi\gamma^2}\Biggr)^{s_0^d}
  \frac{\Gamma(-s_0^d)}{\gamma} \nonumber\\
  &=&\Biggl(\frac{\mu^2}{8\pi\gamma^2}\Biggr)^3
  \prod_{d=1}^r\frac{\Gamma(-s_0^d)}{\Gamma(-s_3^d)}
  A_3(\gamma\alpha^a,\gamma\alpha^b,\gamma\alpha^c)~,
\end{eqnarray}
where $\gamma s_0^d=2\lambda^d\cdot Q/(\alpha^d)^2$ and $s_3^d
=s_0^d-\delta^{ad}-\delta^{bd}-\delta^{cd}$. 
The immediate consequece of (\ref{0ptf}) is that $C_3(\gamma\alpha^a,
\gamma\alpha^b,\gamma\alpha^c)$ is independent of the combinations of 
simple roots. Since the dependences on the simple roots only arise from 
the $\Gamma$-functions in (\ref{0ptf}), we can therefore find the normalized 
three-point functions for Toda vertices 
${\rm e}^{\gamma\alpha^a\cdot\varphi}$ ($a=1,\cdots,r$) in closed form 
as 
\begin{eqnarray}
  \label{norm3ptf}
  \langle{\rm e}^{\gamma\alpha^a\cdot\varphi(z_1)}
  {\rm e}^{\gamma\alpha^b\cdot\varphi(z_2)}
  {\rm e}^{\gamma\alpha^c\cdot\varphi(z_3)}\rangle
  &=&\frac{1}{A_0}G_3(z_1,z_2,z_3|\gamma\alpha^a,\gamma\alpha^b,
  \gamma\alpha^c) \nonumber\\
  &=&-\frac{\Biggl(\frac{\mu^2}{8\pi\gamma^2}\Biggr)^{-3}}
    {|z_{12}|^2|z_{23}|^2|z_{31}|^2}
  \times\cases{s_0^as_0^bs_0^c
    &($a\ne b\ne c$)\cr
   s_0^as_0^b(s_0^b-1)
   &($a\ne b=c$)\cr
   s_0^a(s_0^a-1)(s_0^a-2)
   &($a=b=c$)}
\end{eqnarray}

The two-point couplings can also be related to the partition function
for Toda vertices ${\rm e}^{\gamma\alpha^a\cdot\varphi}$ ($a=1,\cdots,r$). 
Using (\ref{2-3}) and (\ref{0ptf}), we find 
\begin{eqnarray}
  \label{0-2}
  A_0=\Biggl(\frac{\mu^2}{8\pi\gamma^2}\Biggr)^{2}
  \frac{A_2(\gamma\alpha^a)}{s_0^a(s_0^a-1)} ~.
\end{eqnarray}
This gives the normalized two-point functions as
\begin{eqnarray}
  \label{norm2ptf}
  \langle{\rm e}^{\gamma\alpha^a\cdot\varphi(z_1)}
  {\rm e}^{\gamma\alpha^a\cdot\varphi(z_2)}\rangle
  &=&\frac{1}{A_0}G_2(z_1,z_2|\gamma\alpha^a,\gamma\alpha^a) \nonumber\\
  &=&\Biggl(\frac{\mu^2}{8\pi\gamma^2}\Biggr)^{-2}
  \frac{s_0^a(s_0^a-1)}{|z_{12}|^{4}} ~.
\end{eqnarray}
Combining (\ref{0ptf}) and (\ref{0-2}), we obtain the normalization free ratio among 
the couplings
\begin{eqnarray}
  \label{ratio}
  \frac{(A_3(\gamma\alpha^a,\gamma\alpha^b,\gamma\alpha^c))^2A_0}
  {A_2(\gamma\alpha^a)A_2(\gamma\alpha^b)A_2(\gamma\alpha^c)}
  =\cases{\frac{s_0^as_0^bs_0^c}{(s_0^a-1)(s_0^b-1)(s_0^c-1)}
   &($a\ne b\ne c$)\cr
   \frac{s_0^a}{s_0^a-1}
   &($a\ne b=c$)\cr
   \frac{(s_0^a-2)^2}{s_0^a(s_0^a-1)}
   &($a=b=c$)}
\end{eqnarray}
This may be useful in comparing the results obtained via different 
schemes with ours. 

We finally note that the zero- and two-point fucntions obtained above 
are also consistent with the general relations
\begin{eqnarray}
  \label{a0-a3}
  \Biggl(\frac{\partial}{\partial\mu^2}\Biggr)^3A_0
  &=&-\frac{1}{V_0}\Biggl(\frac{1}{8\pi\gamma^2}\Biggr)^3\sum_{a,b,c}
  \int d^2z_1d^2z_2d^2z_3G_3(z_1,z_2,z_3|\gamma\alpha^a,\gamma\alpha^b,
  \gamma\alpha^c) ~,\nonumber \\
  \frac{\partial}{\partial\mu^2}G_2(z_1,z_2|\beta,\beta)&=&
  -\frac{1}{V_2}\frac{1}{8\pi\gamma^2}\sum_a\int d^2z_3
  G_3(z_1,z_2,z_3|\beta,\beta,\gamma\alpha^a) ~,
\end{eqnarray}
where $V_0$ denotes the infinite volume of the M\"obius transformations 
given by
\begin{eqnarray}
  \label{V0}
  V_0=\int\frac{d^2z_1d^2z_2d^2z_3}{|z_{12}|^2|z_{23}|^2|z_{31}|^2} ~.
\end{eqnarray}

We have found a kind of universality 
among the three-point coupling $C_3$ for vertex functions associated 
to the simple roots. The dependences on the simple roots appears 
only in the $\Gamma$-functions which arise in integrating over the 
zero-modes. This enables us to determine all the three-point vertex 
functions associated to simple roots up to an overall normarization. 
We may stress that all these results can be obtained essentially 
from the properties (\ref{i'str}) without knowing the explicit 
form of the three-point couplings. 

\vskip .3cm
This work is supported in part by the Grand-in-Aid for Scientific 
Research from the Ministry of Education, Science and 
Culture (No. 08640348). 
 
\eject

\vfill
\newpage
\begin{thebibliography}{99}
\bibitem{goulian91} M. Goulian and M. Li, Phys. Rev. Lett. {\bf 66} 
(1991) 2051.

\bibitem{gupta} A. Gupta, S. P. Trivedi and M. B. Wise, Nucl. Phys. 
{\bf B340} (1990) 475.

\bibitem{francesco} P. Di Francesco and D. Kutasov, Phys. Lett. {\bf B261} 
(1991) 385;
\newline
Y. Kitazawa, Phys. Lett. {\bf B265} (1991) 262.

\bibitem{sakai}  N. Sakai and Y. Tanii, Prog. Theor. Phys. 
{\bf 86} (1991) 547.

\bibitem{aoki} K. Aoki and E. D'Hoker, Mod. Phys. Lett. {\bf A7} (1992) 235.

\bibitem{dorn94} H. Dorn and H. -J. Otto, Phys. Lett. {\bf B291} (1992) 39;
Nucl. Phys. {\bf B429} (1994) 375.

\bibitem{zamolodchikov} A. B. Zamolodchikov and Al. B. Zamolodchikov, 
Nucl. Phys. {\bf B477} (1996) 577.

\bibitem{wsymm} A. B. Zamolodchikov, Theor. Mat. Phys. {\bf 65} (1986) 1205;
\newline
V. A. Fateev and A. B. Zamolodchikov, Nucl. Phys. {\bf B280} (1987) 644;
\newline
V. A. Fateev and S. L. Luk'yanov, Int. J. Mod. Phys. {\bf A3} (1988)
507.

\bibitem{bilal} A. Bilal and J. -L. Gervais, Phys. Lett. 
{\bf B206} (1988) 412;
Nucl. Phys. {\bf B314} (1989) 646; {\bf B318} (1989) 579; 
{\bf B326} (1989) 222;
\newline
O. Babelon, Phys. Lett. {\bf B215} (1988) 523;

\bibitem{mansfield} P. Mansfield and B. Spence,  Nucl. Phys. 
{\bf B362} (1991) 294;
\bibitem{ddk} F. David, Mod. Phys. Lett. {\bf A3} (1988) 1651;
\newline
J. Distler and H. kawai, Nucl. Phys. {\bf B321} (1989) 509.

\bibitem{hollowood89} T. Hollowood and P. Mansfield, Phys. Lett. 
{\bf B226} (1989) 73.

\bibitem{mansfield83} P. Mansfield, Nucl. Phys. {\bf B222} (1983) 419.

\bibitem{dotsenko} Vl. S. Dotsenko and V. A. Fateev, Nucl. Phys. 
{\bf B240}, 312 (1984); {\bf B251} (1985) 691.

\end{thebibliography}
\end{document}